\documentclass[12pt]{article}
\usepackage[affil-it]{authblk}
\usepackage{amsmath,amssymb}
\usepackage{graphicx}
\usepackage{psfrag}
\usepackage{float}

\numberwithin{equation}{section}

\begin{document}

\title{Interacting thin shells in  the interior of a Reissner-Nordstr\o m black hole}

\author{Alexey L. Smirnov%
\thanks{Electronic address: \texttt{smirnov@ms2.inr.ac.ru}}}

\affil{Institute for Nuclear Research of the Russian Academy of Sciences 60-th October Anniversary Prosp., 7a, 117312, Moscow, Russia}

\maketitle

\begin{abstract}

In this paper we consider some applications of the extended Dray-t'Hooft-Redmount relation, considered earlier in \cite{Neronov0,Maeda,BS}. In particular, using this relation, we study geometries of interacting thin shells near the future Cauchy horizon
of a Reissner-Nordstr\o m black hole.

\end{abstract}

\section{Introduction}
\indent
\label{intro}

Two interacting  ultra-relativistic flows of matter in General Relativity  can be successfully described
by two interacting thin null shells. At the quantitative level, for spherically-symmetric spacetimes, this interaction 
is described by the so-called
Dray-t'Hooft-Redmount (DTR) relation \cite{Redmount,DT}. This expression relates geometries before and after interaction.

On the other hand, there are situations where one or both shells should be considered as time-like ones. This requires
an extension of the DTR relation. Such an extension for mixed systems of thin time-like and null shells was considered
in \cite{Neronov0,Maeda,BS}. In this case however, the parameter space is enlarged and complications arise.
The new relation will include shells velocities and additional geometrical parameters. 
Therefore, to see the physics behind the extended DTR, we should consider examples.

One particular example, where the original DTR relation was successfully exploited is
the mass inflation phenomenon \cite{Israel6}.
The mass inflation is a violent increase of the mass parameter near the future Cauchy horizon of 
a charged black hole.
Initially, such a model was constructed by Israel and Poisson in \cite{Israel6}. The key component of the model
is an ingoing null shell propagating infinitesimally close to the future Cauchy horizon. Another null shell crosses
the horizon and triggers the mass inflation. 
The ingoing shell's energy blows up for the observers crossing the shell. 
Thus such a shell can represent blue-shifted radiation near the
horizon. Then the mass parameter tends to infinity in the region between the shells after their interaction. 
In this way, an effect similar to the mass inflation is obtained.
 
One can consider a generalisation of the Israel-Poisson model where the null shell 
crossing the Cauchy horizon is replaced by a collapsing time-like shell.
In this case, time-like shell represents an ensemble of infalling observers. 
Then using the extended DTR, it is possible to explore the question about the fate of observers crossing 
the future Cauchy horizon of the charged black hole when backreaction is taken into account. 
We point out that similar problem was considered in \cite{Ori} using perturbative analysis near the future Cauchy horizon.
The authors of \cite{Ori} conclude that observers can experience finite tidal effects near the Cauchy horizon.

%One unfortunate feature of the crossing shells is that all interactions details near 
%the crossing point are swept under the rug. Therefore we are forced to do some ad hoc assumptions.

In our model, observers ``burn down'' on the Cauchy horizon - 
the time-like shell effectively turns into a null shell after crossing with the null shell propagated 
along the Cauchy horizon. Moreover, the mass inflation phenomenon is still present in the model.

Another extension considered in the paper takes the original Israel-Poisson model as its essential part. 
In this case the timelike shell enters the region behind the Cauchy horizon where the mass inflation already takes place.
We will be interested in the geometry of the region between timelike an ingoing null shells. It appears that apart from 
the mass inflation scenario in this region there is a choice of parameters which leads to the RN geometry with finite mass.

The paper is organized as follows. Section \ref{sec:0} contains some previously known technical results.
In particular, definitions of ${\cal R}$-, ${\cal T}$-regions are presented, since they are extensively used throughout 
the paper. Next, the theory of thin shells is elucidated. Finally in the section, the extended Dray-t'Hooft-Redmount relation is
obtained.  In Section \ref{sec:1} two extensions of the Israel-Poisson model are constructed and studied.  
Section \ref{sec:2} contains summary and conclusions.
\section{The toolbox: ${\cal R}$-, ${\cal T}$-regions, thin shells and their crossings} 
\label{sec:0}
\subsection{General structure of spherically symmetric spacetime}
\label{ssec:1}
Any  spherically symmetric spacetime $M$ can be considered as a collection of the so-called ${\cal T}$- and ${\cal R}$-regions 
separated by apparent horizons. 
Since  notions of the  ${\cal T}$- and ${\cal R}$-regions will be extensively used throughout  the paper, 
let us recall their definitions \cite[Sec. 2.4.2]{Frolov98}. 
Spherically symmetric space-time metric can be written in the form 
\begin{equation}
\label{ssm}
ds^2=h_{ij}dx^idx^j - r^2(x^0, x^1)(d\theta^2+ \sin^2\theta d\phi^2),\quad i,j=0,1
\end{equation}
and can be locally described by only two 
functions  \cite{Shells13,BS}. The first one is the radius $r(x^0,x^1)$ which is defined in such a way that the area of 
the sphere equals $4\pi r^2$. The second one is the square of the vector normal to the surfaces of constant 
radius $\Delta(x^0,x^1)=h^{ij}\partial_ir\partial_jr$.

We say that a given point $p \in M$ belongs to a ${\cal R}$-region if 
the $r=\mbox{const}$, $\theta=\mbox{const}$,  $\phi=\mbox{const}$ world line is timelike ($\Delta<0$) 
in the neighbourhood of that point. If this world line is spacelike ($\Delta>0$), we say that the given point belongs to  
a ${\cal T}$-region.

Let us fix the remained gauge freedom in \eqref{ssm} by a transformation to coordinates $(t,y)$
where $h_{ij}$ has diagonal form.
Then it is evident that in the ${\cal R}$-region we cannot have $\partial r/\partial y =0$. 
So the sign of partial derivative of the radius is 
an invariant. Thus, we have either $\partial r/\partial y >0$ which is called the ${\cal R_+}$-region, or  
$\partial r/\partial y < 0$, called the ${\cal R_-}$-region.

Similarly, for a ${\cal T}$-region we can never have $\partial r/\partial t =0$. 
Thus, there may be regions of inevitable expansion 
with $\partial r/\partial t >0$, and  regions of inevitable contraction with $\partial r/\partial t <0$. 
The former are called ${\cal T_{+}}$-regions, while the latter - ${\cal T_{-}}$-regions.

The null surfaces ${\cal A}$  of constant radius with $\Delta=0$ define either an apparent horizon 
(a marginally trapped surface) or a anti-trapped surface. 
${\cal A}$ serve as boundaries between ${\cal R}$- and ${\cal T}$-regions.

\begin{figure}[H]
  \psfragscanon
  \psfrag{a}[c][c][0.9][0]{$(a)$}
  \psfrag{b}[c][c][0.9][0]{$(b)$}
  \psfrag{r0}[c][c][0.7][90]{$r=0$}
  \psfrag{rp1}[c][c][0.7][45]{$r=r_+$}
  \psfrag{rp3}[c][c][0.7][-45]{$r=r_+$}
  \psfrag{rp4}[c][c][0.7][-45]{$r=r_+$}
  \psfrag{rm1}[c][c][0.7][45]{$r=r_-$}
  \psfrag{rm2}[c][c][0.7][-45]{$r=r_-$}
  \psfrag{rp2}[c][c][0.7][45]{$r=r_+$}
  \psfrag{Ip1}[c][c][0.7][0]{${\cal I}^{+}$}
  \psfrag{Ip2}[c][c][0.7][0]{${\cal I}^{+'}$}
  \psfrag{Im1}[c][c][0.7][0]{${\cal I}^{-}$}
  \psfrag{Im2}[c][c][0.7][0]{${\cal I}^{-'}$}
  \psfrag{ip1}[c][c][0.7][0]{$i^{+}$}
  \psfrag{ip2}[c][c][0.7][0]{$i^{+'}$}
  \psfrag{im1}[c][c][0.7][0]{$i^{-}$}
  \psfrag{im2}[c][c][0.7][0]{$i^{-'}$}
  \psfrag{i01}[c][c][0.7][0]{$i^{0}$}
  \psfrag{i02}[c][c][0.7][0]{$i^{0'}$}
  \psfrag{Rp}[c][c][0.8][0]{${\cal R_+}$}
  \psfrag{Rm}[c][c][0.8][0]{${\cal R_-}$}
  \psfrag{Tp}[c][c][0.8][0]{${\cal T_+}$}
  \psfrag{Tm}[c][c][0.8][0]{${\cal T_-}$}
  \includegraphics[width=0.7\textwidth]{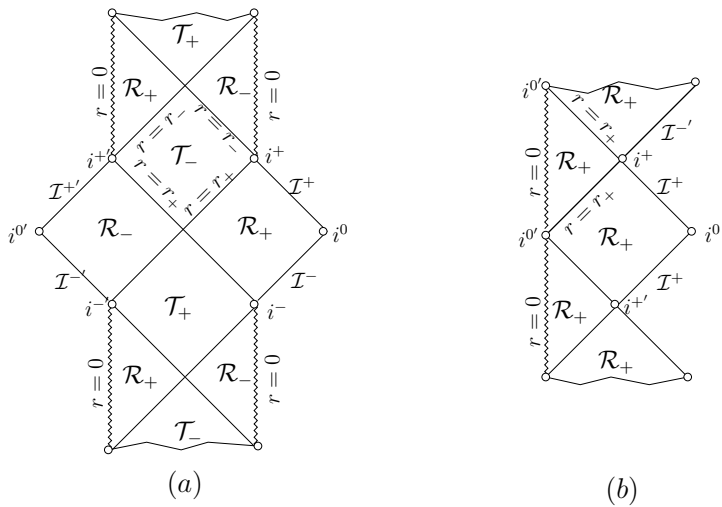}
  \caption{Parts (cut along  broken lines) of maximal analytic extensions for RN spacetimes: 
  (a) Carter-Penrose diagram for a non-extremal RN. (b) Carter-Penrose diagram for an extremal RN.}
\label{fig:0}      
\end{figure}

In many cases, however, ${\cal R}$- and ${\cal T}$-regions can be recognised more easily when \eqref{ssm}
is written in double null coordinates \cite{Dafermos05}
\begin{equation}
\label{ssm1}
ds^2=2h(u,v)dudv - r^2(u,v)(d\theta^2+ \sin^2\theta d\phi^2),
\end{equation}
Metric \eqref{ssm1} is future oriented in the standard way so that $(u,v)$ are both increasing towards the future.
Factoring out the $SO(3)$ group action, we have for a point $q \in M/SO(3)$
\begin{eqnarray}
\label{rtregs}
q &\in& {\cal R_+}\quad\mbox{iff}\quad\partial_u r<0,\ \partial_v r>0\ \mbox{at}\ q,\nonumber\\
q &\in& {\cal R_-}\quad\mbox{iff}\quad\partial_u r>0,\ \partial_v r<0\ \mbox{at}\ q,\nonumber\\
q &\in& {\cal T_+}\quad\mbox{iff}\quad\partial_u r>0,\ \partial_v r>0\ \mbox{at}\ q,\nonumber\\
q &\in& {\cal T_-}\quad\mbox{iff}\quad\partial_u r<0,\ \partial_v r<0\ \mbox{at}\ q,\nonumber\\
q &\in& {\cal A}\quad\mbox{ if}\quad\partial_u r=0\quad \mbox{or}\quad \partial_v r=0\ \mbox{at}\ q.
\end{eqnarray}
This representation is especially useful in the case of (electro)vacuum spacetimes, since  Carter-Penrose diagrams 
are graphical representations of metrics of the form \eqref{ssm1}. In particular, for RN spacetimes,  
positions of ${\cal R}$- and ${\cal T}$-regions in Carter-Penrose diagrams are shown in Fig.~\ref{fig:0}.

\subsection{Thin shells}
\label{ssec:2}
In the present paper, our main tool to model matter with backreaction is the theory of thin shells. 
The general mathematical theory of thin shells was introduced in \cite{Israel13,Shells13}. 
Thin shells are $C^1$-singular objects, since a non-zero amount of energy is confined in vanishing volume.
In other words, while the metric coefficients are continuous across a shell's hypersurface, their derivatives undergo a jump. 
This jump is governed by the so-called Israel equations. In the case of time-like hypersurface the Israel equations 
connect the surface energy-momentum tensor $S_i^j$ of the shell to the jump in the extrinsic curvature 
tensor $K_i^j$ describing embedding of the shell's hypersurface into the geometry on the corresponding side of the shell.  

Let us obtain these equations in the case of spherical symmetry.
First of all, we introduce the Gaussian normal coordinates associated with the shell. 
\begin{eqnarray}
\label{gm}
ds^2 &=& -dn^2 + \gamma_{ij}(n,x)dx^idx^j,\quad i,j=0,2,3\nonumber\\
     &=& \gamma_{00}(t,n)dt^2 - dn^2 - R^2(t,n)(d\theta^2+ \sin^2\theta d\phi^2).
\end{eqnarray}
The shell is situated at $n = 0$ and  $n$ is the spatial coordinate normal to the shell ($n<0$ inside and $n>0$ outside). 
The extrinsic curvature tensor is defined the as $K_i^j=-\frac{1}{2}\gamma_{ij,n}$, here comma denotes a partial derivative.
The surface energy-momentum tensor of the shell is defined by $T_{\mu}^{\nu}=S_{\mu}^{\nu}\delta(n)+~\dots$, 
where dots define nonsingular terms. Since the metric coefficients of \eqref{gm}
are continuous on the shell, so if some of their first derivatives undergo jumps at the shell position, 
the corresponding second derivatives contained in the Einstein equations, will exhibit the $\delta$-function behaviour. 
Integration across the shell yields $S_n^n=0$ and $S_i^n=0$, that can be viewed as the shell definition, and the desired Israel 
equations
\begin{equation}
\label{gie}
-[K_i^j]+\delta_i^j[K]=8\pi S_i^j,
\end{equation}
here square brackets stand for the jump.
Additionally, from the Bianci identities follows the so-called continuity equation for $S_i^j$:
\begin{equation}
\label{gqe}
S_{i|j}^j+[T_i^n]=0,
\end{equation} 
where the vertical line denotes covariant differentiation with respect to the metric on the shell $\gamma_{ij}(0,x)$.
Because of the spherical symmetry $K_2^2=~K_3^3$, $S_2^2=~S_3^3$, \eqref{gqe} is reduced to
%\begin{subequations}
\begin{eqnarray}
\label{ie}
[K_2^2] &=& 4\pi S_0^0,\nonumber\\
{[K_0^0]}+{[K_2^2]} &=& 8\pi S_2^2,
\end{eqnarray} 
%\end{subequations}
Let us introduce the proper time $\tau$ for the observers sitting on the shell, by $d\tau=\gamma_{00}(t,0)dt$ and denote
$r(\tau)=R(t,0)$. Using $\Delta$-invariant from the previous section we can easily calculate $K_2^2$. Indeed, in gaussian 
normal coordinates $\Delta=\gamma^{00}R,_{t}^2-R,_{n}^2$ and thus
\begin{equation}
K_2^2=-\frac{R,_{n}}{R}=-\frac{\sigma}{r}\sqrt{\dot r^2-\Delta},
\end{equation} 
here overdot denotes differentiation with respect to $\tau$ and $\sigma$ is the sign of the outward normal to the shell.
It follows form definition of ${\cal R_{\pm}}$-regions that $\sigma=+1$ in the ${\cal R_+}$-region and  $\sigma=-1$ 
in the ${\cal R_-}$-region. {\it The parameter $\sigma$ cannot change its sign in ${\cal R}$-regions on the equations of motion. 
In fact, $\sigma$ can only be changed in ${\cal T_{\pm}}$-regions}. In particular, for the RN spacetime, 
this property allows charged shell reach both the internal ${\cal R_+}$-region 
and the internal ${\cal R_-}$-region \cite{Boulware}.

The calculation of $K_0^0$ is more involved  and presented in \cite{Shells13}. Here we show only the result:
\begin{equation}
K_0^0= -\frac{\sigma}{\sqrt{\dot r^2-\Delta}}\left(\ddot r +\frac{1+\Delta}{2r}-4\pi r T_n^n\right)
\end{equation}  
Also, the continuity equation \eqref{gqe} can be rewritten in the form
\begin{equation}
\label{qe}
\dot S_0^0 + \frac{2\dot r}{r}(S_0^0-S_2^2)+[T_0^n]=0
\end{equation} 

Equations \eqref{ie}, \eqref{qe} define evolution of the shell completely. However, for the rest of the paper essential 
for us will be the $(_0^0)$-equation in \eqref{ie}. Moreover,
we are interested in the case when the shell is a dust shell and the spacetime off the shell is a RN spacetime. 
Then $\Delta=-F$, where $F$ is the usual coefficient in the static RN metric. 
Also, since $S_2^2=0$ for the dust, it follows from \eqref{qe} that $S_0^0=M/4\pi r^2$, where
$M=\mbox{const}>0$ is the rest mass of the shell. Thus $(_0^0)$-equation in \eqref{ie} has the following form:
\begin{equation}
\label{eqmo}
\sigma_{\rm in}\sqrt{\dot r^2+F_{\rm in}}-\sigma_{\rm out}\sqrt{\dot r^2+F_{\rm out}}=\frac{M}{r}.
\end{equation}  

For light-like (null) shells the Israel equations \eqref{gie} degenerate into the trivial identity 0=0, since
the shell's normal is tangential to the shell's hypersurface.
Thus the junction equation in this case should be derived separately \cite{Shells13}. However, we will use
an alternative approach for (electro)vacuum spacetime with a null uncharged shell. 
It uses the Vaidya metric which describes the gravitational 
field of an unidirectional radial flow of null uncharged dust \cite{Vadya1}.
Then the (electro)vacuum spacetime with the null shell is a special case of the Vaidya metric where the mass function 
is proportional to a step function. However, the original Vaidya metric is incomplete. 

Instead, it is preferable to use
the Vaidya metric in double null coordinates \cite{Vadya5}.  
For the flow along the $u$-direction (i. e. $v=v_0$=const), the only non-zero component of the energy-momentum 
tensor $T_{vv} = f(u,v)/8\pi$. The mass function is $m(v)= m_{\rm in} + \delta m \theta(v-v_0)$, where $\delta m \equiv m_{\rm out} - m_{\rm in}$ is the total mass of the shell.
Then the Einstein equations are reduced to the following set of equations
%\begin{subequations}
\begin{eqnarray}
  h &=& 2B(v) \partial_u r\label{vdn1},\\
  \partial_v r &=& -B(v)\left(1-\frac{m(v)}{r}\right),\label{vdn2}\\
  f &=& -4\frac{B(v)}{r^2}\partial_v m(v).\label{vdn3}
\end{eqnarray}
%\end{subequations}
Here $B(v)$ is an arbitrary function of $v$ only. Note, the metrics is explicitly continuous across the shell.
In fact, $B$ defines the scale for $v$ which is required for a proper definition of $m$.
Then it follows from \eqref{vdn3}, that $vv$-component of the shell's energy-momentum is 
\begin{equation}
\label{tvv}
T_{vv}=-\frac{B(v)\delta m}{2\pi r^2} \delta(v-v_0),
\end{equation} 
If the weak energy condition is satisfied and $\delta m>0$, then $B(v) \leq 0$. 
An extension to the case of charged null dust is also possible \cite{Vadya2}. 

\subsection{Crossing thin shells}
\label{ssec:3}
\begin{figure}[t]
  \begin{center}
    \psfragscanon
    \psfrag{uv1}[c][c][0.9][0]{$(u_1,v_1)$}
    \psfrag{uv2}[c][c][0.9][0]{$(u_2,v_2)$}
    \psfrag{uv3}[c][c][0.9][0]{$(u_3,v_3)$}
    \psfrag{uv4}[c][c][0.9][0]{$(u_4,v_4)$}
    \psfrag{uv}[c][c][0.9][0]{$(u,v)$}
    \includegraphics[width=0.5\textwidth]{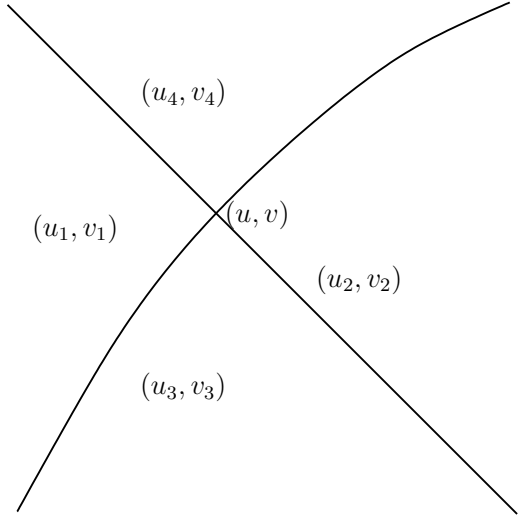}
    \caption{Schematic representation of two crossing shells. Coordinates $(u_1,v_1)$ cover the interior
      of the null shell before crossing, $(u_2,v_2)$ cover the exterior of the time-like shell, and $(u_3,v_3)$, 
      $(u_4,v_4)$ cover the region between shells before and after crossing correspondingly.}
    \label{fig:1}
  \end{center}    
\end{figure}

In this section we present relevant results from the theory of colliding (crossing) 
spherically-symmetric shells \cite{Neronov0,BS}. 
In the following, the term 'crossing' will be used to emphasize that the worldtubes of shells cross each other.  
In the neighbourhood of the shells interaction point the spacetime is $C^0$-regular.
We are interested in the case when one light-like shell crosses one time-like shell.
Timelike and null shells collapse before and after crossing.
Informal picture for this crossing is shown at the figure Fig.~\ref{fig:1}.
Each sector $i$ where $i=1,2,3,4$ is described by a vacuum spherically-symmetric metric and have 
its own set of double null coordinates $(u_i,v_i)$. Written in these coordinates metrics have the form 
\begin{equation}
\label{bds}
ds^2 = 2h_i(u_i,v_i) du_i dv_i - r_i^2(u_i,v_i) d\Omega^2.
\end{equation}
These metrics are discontinuous on the shells. However the whole spacetime can be described by a $C^0$-metric. Thereby,
we can introduce double null coordinates $(u,v)$ covering the vicinity of the crossing point and for these coordinates
there exists a continuous metric
\begin{equation}
  \label{cds}
  ds^2 = 2h(u,v) du dv - r^2(u,v)d\Omega^2.
\end{equation}
For every sector $i$, coordinates  $(u_i,v_i)$ are related with $(u,v)$ as follows
\begin{eqnarray}
  u_i &=& u_i(u)\nonumber \\ v_i &=& v_i(v)\nonumber.
\end{eqnarray}
Transformations of the metric coefficients in \eqref{bds} to \eqref{cds} yields
  \begin{eqnarray}
    \label{rel} 
    h(u,v) &=& h_i(u_i(u),v_i(v)) u_i^{ \prime}(u) v_i^{\prime}(v)\nonumber, \\ 
    r(u,v) &=& r_i(u_i(u),v_i(v)).
  \end{eqnarray}
Thus, if we sit on the shell between regions  $i$ and $(i+1)$ then 
\begin{eqnarray}
  \label{con1}
  h_i(u_i,v_i) u_i^{\prime} v_i^{\prime} |_{\rm shell} &=&
  h_{i+1}(u_{i+1},v_{i+1}) u_{i+1}^{\prime} v_{i+1}^{\prime} |_{\rm shell},\\
  \label{con2}
  r_i(u_i,v_i)|_{\rm shell} &=& r_{i+1}(u_{i+1},v_{i+1})|_{\rm shell}.
\end{eqnarray} 
In general, trajectories of shells can be written as follows 
\begin{eqnarray}
  \label{sur1}
  \Sigma_i (u_i,v_i) &=& 0,\\
  \label{sur2}
  \Sigma_{i+1} (u_{i+1},v_{i+1}) &=& 0
\end{eqnarray}
for sectors $i$, $i+1$ respectively.
Let's differentiate  \eqref{con2} by $u$ along the shell. The result is
\begin{equation}
\label{diff}
\left[ \partial_{u_i} r_i + 
\partial_{v_i} r_i \left(\frac{dv_i}{du_i}\right)_{\Sigma_i} \right] 
u_i^{\prime} = \left[ \partial_{u_{i+1}} r_{i+1} +
\partial_{v_{i+1}} r_{i+1} \left(\frac{dv_{i+1}}{du_{i+1}}\right)_{\Sigma_{i+1}}
\right] u_{i+1}^{\prime}
\end{equation}
The symbol  $( )_{\Sigma}$ means that we should differentiate the implicit function $u_i(v_i)$,  $u_{i+1}(v_{i+1})$ using
\eqref{sur1}, \eqref{sur2} respectively. 
Expression similar to \eqref{diff} can be obtained also with respect to the coordinate $v$. The procedure is now clear.
Recursively using \eqref{con1}, \eqref{diff} at the crossing point, we can eliminate $u_i^{\prime}$ (or $v_i^{\prime}$) and obtain 
extension of the Dray-'tHooft-Redmount relation.

In our case, metrics \eqref{bds} is derived from the standard vacuum spherically-symmetric metrics
$$
\ ds^2 = F_i dt_i^2 - F_i^{-1} dr_i^2-r^2_id\Omega^2,
$$
by coordinate transformations
\begin{eqnarray}
\label{dnt}
u_i &=& t_i- \sigma_i r^*_i,\\
v_i &=& t_i+ \sigma_i r^*_i\nonumber
\end{eqnarray}
Here $r^*_i$ is the 'tortoise' coordinate and $\sigma_i = +1$, if  $r^*_i$ increases along the $r_i$ axis 
and $\sigma_i = -1$ if $r^*_i$ increases to opposite direction of the $r_i$ axis.  When \eqref{dnt} is restricted 
to the time-like shell, $\sigma_i$ plays the same role as in the previous section i.e. it can be viewed as the sign of the
outer normal to the shell. Transformations \eqref{dnt} give
$$ 
h_i=F_i 
$$
in \eqref{bds}  and
\begin{eqnarray}
\label{nder}
\partial_{u_i} r_i &=& -\frac{\sigma_i F_i}{2},\\
\partial_{v_i} r_i &=& \frac{\sigma_i F_i}{2}\nonumber.
\end{eqnarray}
Sectors 2 and 3 are joined along the light-like shell. Then if we align $u_1$, $u_3$ along the shell 
\begin{eqnarray}
\label{s13}
\Sigma_1 &=& v_2-C_2,\\
\Sigma_3 &=& v_3-C_3,\nonumber
\end{eqnarray}
where $C_1$, $C_3$ are some constants.
From \eqref{con1}, \eqref{diff} by using \eqref{nder} and \eqref{s13} we obtain transition formulae between sectors 2 and 3
\begin{eqnarray}
\label{v13}\sigma_2 F_2 u_2^{\prime} &=& \sigma_3 F_3 u_3^{\prime},\\
\label{u13}\sigma_2 v_2^{\prime} &=& \sigma_3 v_3^{\prime}
\end{eqnarray}
Similarly, transition formulae between sectors 1 and 4 yields 
\begin{eqnarray}
\label{v24} \sigma_1 F_1 u_1^{\prime} &=& \sigma_4 F_4 u_4^{\prime},\\
\label{u24} \sigma_1 v_1^{\prime} &=& \sigma_4 v_4^{\prime}
\end{eqnarray}

Sectors 1,3 and 2,4 are joined along the time-like shell. Again, to obtain transition formulae we should calculate coefficients
in \eqref{diff}. For definiteness, since calculations are similar in each sector, we will work in sector 1. Trajectory of the
shell is given by
$$
\Sigma_1 = r_1 -r(t_1),
$$
the $C^0$-function $r(t_1)$ and away form the crossing point it must be defined by the equation of motion of the shell. 
Then on $\Sigma_1$
$$
 \partial_{u_1} r_1 + \partial_{v_1} r_1 \left(\frac{dv_1}{du_1}\right)_{\Sigma_1}=\frac{F_1dr/dt_1}{F_1+\sigma_1dr/dt_1}.
$$
Defining the proper time $\tau$ on the shell we have 
$$
\left(\frac{d\tau}{dt_1}\right)^2=F_1-\frac{1}{F_1}\left(\frac{dr}{dt_1}\right)^2\quad \mbox{and} \quad \frac{dr}{dt_1}=\dot r\frac{d\tau}{dt_1},
$$ 
Hence, eventually
\begin{equation}
\label{coeff}
 \partial_{u_1} r_1 + \partial_{v_1} r_1 \left(\frac{dv_1}{du_1}\right)_{\Sigma_1}=\frac{F_1\dot r_b}{\sqrt{\dot r^2_b+F_1}-\sigma_1\dot r_b}.
\end{equation}
Since $\dot r(\tau)$ is discontinuous at the crossing point, we adopted the convention
$$
\dot r_b =\dot r|_{\tau\to\tau^-_c},\quad\dot r_a =\dot r|_{\tau\to\tau^+_c},
$$ 
where $\tau_c$ is the time of crossing.

Using \eqref{coeff} remaining transition formulae can be written as follows
\begin{eqnarray}
\label{v14} u_2^{\prime} &=& \frac{F_4\left( \sqrt{\dot r^2_a +F_2} - 
\sigma_2 \dot r_a\right)}{F_2\left(\sqrt{\dot r^2_a + F_4} - \sigma_4 \dot r_a \right)} u_4^{\prime}, \\
\label{v23} u_1^{\prime} &=& \frac{F_3\left( \sqrt{\dot r^2_b +F_1} - 
\sigma_1 \dot r_b\right)}{F_1\left(\sqrt{\dot r^2_b + F_3} - \sigma_3 \dot r_b \right)} u_3^{\prime}.
\end{eqnarray}

Now successively eliminating $u_i^{\prime}$ from \eqref{v13}, \eqref{v24}, \eqref{v14}, \eqref{v23} 
we obtain desired extension of the Dray-'tHooft-Redmount relation
\begin{eqnarray}
  \label{edtr}
  \sigma_1\sigma_2\left(\sqrt{\dot r^2_b + F_1} - \sigma_1 \dot r_b \right)&\left(\sqrt{\dot r^2_a + F_2} + \sigma_2 \dot r_a \right)&=\nonumber\\
  =\sigma_3\sigma_4\left( \sqrt{\dot r^2_a +F_4} - \sigma_4 \dot r_a\right)&\left( \sqrt{\dot r^2_b +F_3} - \sigma_3 \dot r_b\right)&.
\end{eqnarray}

Note, that factors $\sigma_1\sigma_2$, $\sigma_3\sigma_4$ were missed in the analysis of \cite{Neronov0,BS}. 
They appears to be important as we will show in Section \ref{ssec:5}. 
The conventional Dray-'tHooft-Redmount relation \cite{Redmount,DT} can be obtained by this procedure almost trivially.
Indeed, replace the time-like shell by another null shell. Then \eqref{v14}, \eqref{v23} are replaced by
\begin{equation}
\sigma_3 u_3^{\prime} = \sigma_1 u_1^{\prime},
\end{equation}
\begin{equation}
\sigma_2 u_2^{\prime} = \sigma_4 u_4^{\prime}.
\end{equation} 
Again, eliminating $u_i^{\prime}$ we obtain
\begin{equation}
\label{dtr}
F_1F_2=F_3F_4
\end{equation}
The same procedure is available as for any number of crossing time-like shells
as for any number of crossing null shells. 

In the end of the section let us make some cautionary remarks. In the above construction, an initial value problem
for a system of several crossing shell is not well defined \cite{Klein}. This happens since we do not define physics near 
the interaction point. The extended DTR cannot solve the problem. 
Then, for example, the above procedure is equally well defined whether or not the number of shells 
is conserved during the interaction. 
Even more, equations of state for shells can be changed during the interaction. In the next
section, to avoid these ambiguities, we will assume that the shells interaction is purely gravitational.

\section{Crossing shells near the future RN Cauchy horizon}
\label{sec:1}
Now, having all necessary tools at our disposal, we can study how crossing shells may be applied to modelling
of matter interactions near Cauchy horizons of charged black holes. 
We start with the well-known Israel-Poisson model and then try to extend it.

\subsection{The Israel-Poisson model}
\label{ssec:4}
\begin{figure}[H]
  \begin{center}
 \psfragscanon
 \psfrag{r0}[c][c][0.9][90]{$r=0$}
 \psfrag{Ip}[c][c][0.9][0]{${\cal I}^{+}$}
 \psfrag{Im}[c][c][0.9][0]{${\cal I}^{-}$}
 \psfrag{ip}[c][c][0.9][0]{$i^+$}
 \psfrag{i0}[c][c][0.9][0]{$i^0$}
 \psfrag{Rp}[c][c][0.9][0]{${\cal R_+}$}
 \psfrag{Rm}[c][c][0.9][0]{${\cal R_-}$}
 \psfrag{Tm}[c][c][0.9][0]{${\cal T_-}$}
 \psfrag{A}[c][c][0.8][0]{$\rm 1$}
 \psfrag{B}[c][c][0.8][0]{$\rm 3$}
 \psfrag{C}[c][c][0.8][0]{$\rm 2$}
 \psfrag{D}[c][c][0.8][0]{$\rm 4$}
 \psfrag{rp}[c][c][0.9][0]{$r_{\rm 3+}$}
 \psfrag{rm}[c][c][0.9][0]{$r_{\rm 2-}$}
 \psfrag{eh}[c][c][0.9][0]{$r_{2+}$}
 \psfrag{ch}[c][c][0.9][0]{$r_{3-}$}
 \psfrag{ah}[c][c][0.9][0]{$r_{4-}$}
 \psfrag{u}[c][c][0.8][0]{$u$}
 \psfrag{v}[c][c][0.8][0]{$v$}
  \includegraphics[width=0.6\textwidth]{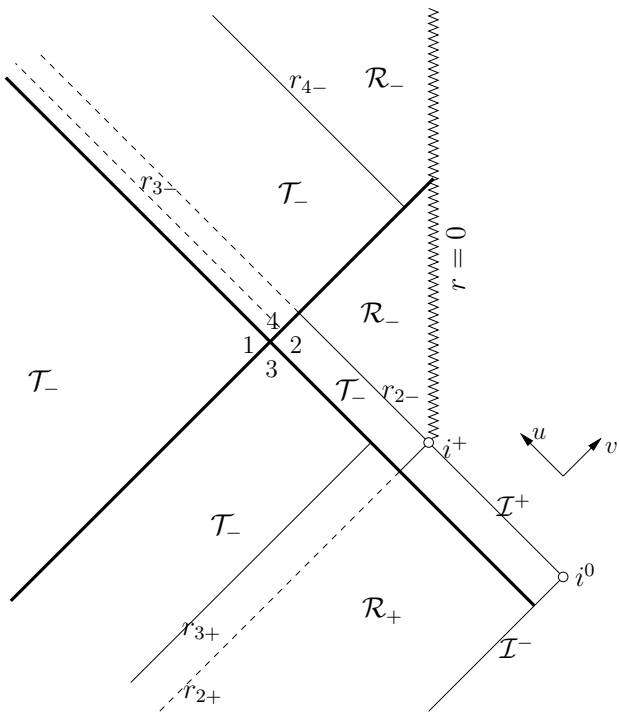}
\caption{Carter-Penrose diagram for the Israel-Poisson model.
  Sector 1 is the interior of the outgoing shell before crossing. It lies entirely in the internal ${\cal T_-}$-region
  of the RN spacetime. Sector 3 is the region between the shells before crossing. Sector 2 is the exterior
  of the ingoing shell. Sector 4 is the mass inflation region. Dashed lines schematically represent
  continuation of horizons between sectors.  The future Cauchy horizon is $r_{3-}$,
  the contracted  inner apparent horizon is $r_{4-}$ which has very small radius and the event horizon of the RN black hole
  is $r_{2+}$. Note, since the total mass of the ingoing shell $m_2-m_3>0$ then $r_{3-}>r_{2-}$.}
\label{fig:2}
  \end{center}
\end{figure}

Two crossing null shells were used by Israel and Poisson as the simplest model to explain 
the mass inflation phenomenon \cite{Israel6}. As it is well known, the mass inflation is a violent increase of 
the mass parameter near the future Cauchy horizon of charged and rotating black holes.
The interior model of RN black hole in \cite{Israel6} (see also \cite[Sec. 14.4.3]{Frolov98}) contains 
two cross-flowing streams of radially moving null dust. 
The ingoing flow models the infalling backscattered radiation which was initially emitted from the 
surface of a collapsing body. The outgoing flow represents the outgoing part of backscattered radiation.    
In the simplest case, these streams can be described by two crossing uncharged light-like shells.
Four sectors of the RN spacetime are glued along these shells. 
The shells crossing point is placed in the region ${\cal T_{-}}$ of Fig.~\ref{fig:0}(a) near the right segment of
the Cauchy horizon. The corresponding Carter-Penrose diagram is presented at Fig.~\ref{fig:2}. 

With this setup,  the mass inflation is a direct consequence of the Dray-'tHooft-Redmount relation \eqref{dtr} 
between mass parameters in sectors 1, 2, 3, 4 in Fig.~\ref{fig:2}. Namely,
\begin{equation}
\label{dtrRN}
\left(1-\frac{m_{\rm 1}}{r_c}+\frac{q^2}{r^2_c}\right)\left(1-\frac{m_{\rm 2}}{r_c}+\frac{q^2}{r^2_c}\right)=\left(1-\frac{m_{\rm 3}}{r_c}+\frac{q^2}{r^2_c}\right)\left(1-\frac{m_{\rm 4}}{r_c}+\frac{q^2}{r^2_c}\right),
\end{equation}
where $r_c$ is the crossing radius. Since shells have vanishing electric charge, the constant $q$ is the same in all sectors.
Let $r_c \to r_{3-}$, then we can see from \eqref{dtrRN} that if the first multiplier in the r.h.s. is very close to zero, 
the second multiplier should be large enough to get finite and non-zero l.h.s. The only way to do that is to 
increase the mass parameter $m_{\rm 4} \to \infty$. 
This is allowed, since the crossing point $r_c$ lies in $\cal T_-$-regions with respect to sectors~1,~2,~3.

That the shell propagating along the inner horizon of the sector 3 mimics the infinitely blue-shifted radiation is better seen
in the Vaidya picture. Let us introduce a family of radially moving observers 
who cross the ingoing shell and calculate its local energy $E$, which the observers measure i.e.
\begin{equation}
\label{le1}
E = 4\pi r^2_0 \int T_{\mu\nu}u^{\mu}u^{\nu}d\tau, 
\end{equation}
where $u^{\mu}$ is the observer's four-velocity and $r_0$ the radius of the shell when the observer crosses it. In double-null
coordinates $(u,v)$, that are regular at the horizon $r_{3-}$, the shell propagates at $v=v_0$ and the observer's trajectory
is given by a regular function $u(v)$. Then using \eqref{tvv}, \eqref{le1} can be written as follows 
\begin{equation}
\label{le2}
E= -\sqrt{\frac{2}{h(v_0)u^{\prime}(v_0)}}B(v_0)\delta m,
\end{equation}
where $\delta m = m_2 - m_3 > 0$.
Further, let us denote by $v_H$ the position of the future Cauchy horizon $r_{3-}$ in $(u,v)$ coordinates and consider behaviour 
of \eqref{le2} when $v_0 \to v_H$. Note, that the metric coefficient $h$ is continuous across the horizon, 
and can be written in the form \eqref{vdn1}. On the other hand, as it follows from \eqref{rtregs}, 
the derivative $\partial_u r \to 0^-$ when  $v_0 \to v_H$. Therefore, we have to set $\lim_{v_0\to v_H}B(v)=-\infty$, 
and then \eqref{le2} is divergent. Therefore, in the Israel-Poisson model, the shell with $v_0 \to v_H$ gives the expected 
flow of infinitely blue-shifted photons near the future Cauchy horizon. 

Thus the Israel-Poisson model gives the effect similar to the mass inflation singularity. 
However, the null shell introduces a singularity of different kind. In fact, this singularity is the property of the Vaidya
spacetimes. The point is that the Cauchy horizon represents a boundary in a Vaidya-RN metrics 
written in Finkelstein-like coordinates.
A smooth extension of this metrics beyond the Cauchy horizon can be constructed only 
if the derivative of the mass function $dm(v_H)/dv=0$ \cite{Israel7,Fayos}. 
Otherwise one obtains a physical $C^1$-singularity due to radiation trapped and piled up along the Cauchy horizon. 
Exactly this happens in the Israel-Poisson model, since $\delta m \neq 0$.  

\subsection{Extensions of the Israel-Poisson model}
\label{ssec:5}
\begin{figure}[H]
  \begin{center}
 \psfragscanon
 \psfrag{r0}[c][c][0.9][90]{$r=0$}
 \psfrag{Ip}[c][c][0.9][0]{${\cal I}^{+}$}
 \psfrag{Im}[c][c][0.9][0]{${\cal I}^{-}$}
 \psfrag{ip}[c][c][0.9][0]{$i^+$}
 \psfrag{i0}[c][c][0.9][0]{$i^0$}
 \psfrag{Rp}[c][c][0.9][0]{${\cal R_+}$}
 \psfrag{Rm}[c][c][0.9][0]{${\cal R_-}$}
 \psfrag{Tm}[c][c][0.9][0]{${\cal T_-}$}
 \psfrag{A}[c][c][0.8][0]{$\rm 1$}
 \psfrag{B}[c][c][0.8][0]{$\rm 3$}
 \psfrag{C}[c][c][0.8][0]{$\rm 2$}
 \psfrag{D}[c][c][0.8][0]{$\rm 4$}
 \psfrag{rp}[c][c][0.9][0]{$r_{\rm 3+}$}
 \psfrag{rm}[c][c][0.9][0]{$r_{\rm 2-}$}
 \psfrag{eh}[c][c][0.9][0]{$r_{2+}$}
 \psfrag{ch}[c][c][0.9][0]{$r_{3-}$}
 \psfrag{ah}[c][c][0.9][0]{$r_{4-}$}
 \psfrag{u}[c][c][0.8][0]{$u$}
 \psfrag{v}[c][c][0.8][0]{$v$}
  \includegraphics[width=0.6\textwidth]{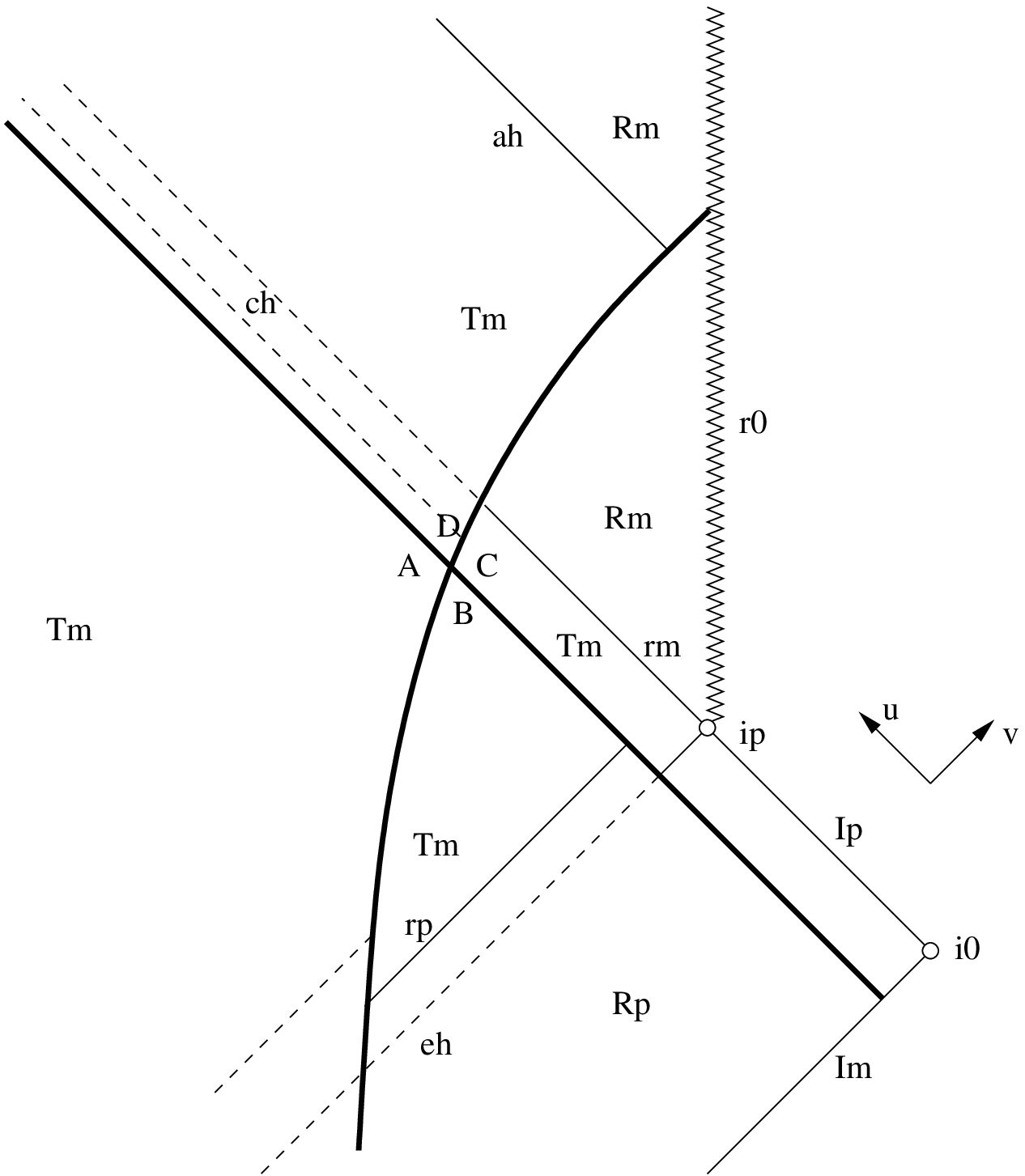}
\caption{}
\label{fig:3}
  \end{center}
\end{figure}

The discussion at the end of the previous section suggests to consider as a separate problem. 
Namely, the fate of observers crossing the future Cauchy horizon with trapped radiation.
However, the above  picture doesn't take into account backreaction of the infalling observer on the geometry.
The simplest approach to overcome this limitation is to represent an ensemble of infalling observers by a time-like shell. 
This leads to the corresponding extension of the Israel-Poisson model using the formalism presented in Section \ref{ssec:3}.   
The extension we are going to consider is depicted at Fig.~\ref{fig:3}. 

Collapsing charged dust shell (infalling observers) with
the rest mass $M>~0$ crosses the future Cauchy horizon of sector~3 with trapped radiation (the null shell). 
Intuitively, one would expect that the mass inflation phenomenon exist also in the setup at Fig.~\ref{fig:3}. 
To prove it one needs to know the behaviour of the dust shell before and after the crossing.

Before the crossing the time-like shell propagate in sectors~1,~3 and is described by \eqref{eqmo}, which can be written as 
$$
\sigma_1\sqrt{\dot r^2+F_1}-\sigma_3\sqrt{\dot r^2+F_3}=\frac{M}{r}.
$$
If the shell collapses from infinity then initially $\sigma_1=1$, $\sigma_3=1$.
However, to reach the future Cauchy horizon of sector~3, $\sigma_3$ must change its sign provided that $r_{3-}\neq r_{1-}$. 
The sign of $\sigma_1$ is preserved. Indeed, if there were no crossing, the shell 
would eventually reach the $\cal R_-$-region in sector~3 
where $\sigma_3=-1$. From the discussion in Section~\ref{ssec:2} and Fig.~\ref{fig:3} it follows that $\sigma_3$ can change 
its sign, since some part of the spacetime in sector~3 lies in $\cal T_-$-region.

On the other hand, since the charged shell is considered, there is also degenerate case when $r_{3-}=r_{1-}$.
Heuristically we would expect that $\sigma_3$ also change its sign in this situation but it is not evident. 

To manage this case, it is instructive to digress and explore more carefully 
how the charge of the collapsing time-like shell affects the horizons of the black hole. 
Later on, to simplify calculations, we consider initially extremal black hole. 

In the uncharged case the outer apparent horizon 
is always increasing and the inner horizon is always decreasing, since $\partial r_+/\partial m >0$, $\partial r_-/\partial m <0$.
This is not the case when collapsing matter is charged, since $\partial r_+/\partial q^2 <0$ and $\partial r_-/\partial q^2 >0$. 

\begin{figure}[H]
  \begin{center}
    \psfragscanon
    \psfrag{r0}[c][c][1.0][0]{$R_0$}
    \psfrag{r}[c][c][1.0][0]{$r$}
    \psfrag{r1}[c][c][1.0][0]{$m_1-M$}
    \psfrag{r2}[c][c][1.0][0]{$m_1+M$}
    \psfrag{rm}[c][c][1.0][0]{$r_{3-}$}
    \psfrag{q1}[c][c][1.0][0]{$\delta q_1^+$}
    \psfrag{q2}[c][c][1.0][0]{$\delta q_2^+$}
    \psfrag{q}[c][c][1.0][0]{$\delta q$}
    \includegraphics[width=0.6\textwidth]{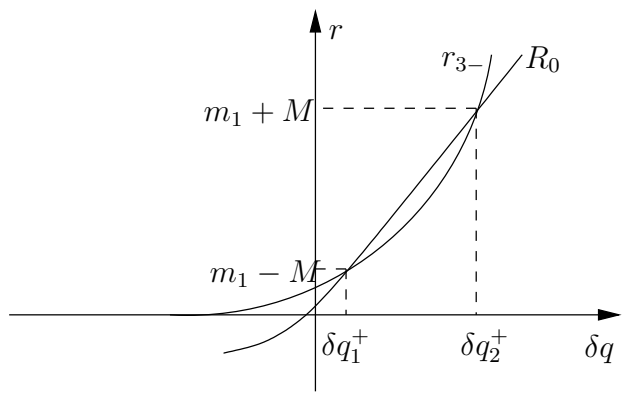}
    \caption{Behaviour of $R_0$, $r_{3-}$ as functions of $\delta q$.}
    \label{fig:4}
  \end{center}    
\end{figure}

Since the black hole is initially the extremal one and the time-like shell is charged, 
the outer apparent horizon is increasing during accretion. However, the inner 
apparent horizon can be either less or greater than the initial extremal horizon. 
Indeed, let $\delta q$ is the charge of the time-like shell and $\delta m = m_3-m_1>0$ its total mass. 
Then behaviour of the inner apparent horizon is governed by inequalities 
\begin{equation}
  \label{iah}(m_1+\delta m) - \sqrt{(m_1+\delta m)^2-(m_1+\delta q)^2} \lesseqgtr m_1.
\end{equation}
The saturation  of \eqref{iah} means that the inner apparent horizon is not changed during collapse i. e.  $r_{3-}= r_{1-}$. 
For given $\delta m$ this happens  when the shell has the either of the following charges
\begin{equation}
\label{qcrit}\delta q^{\pm}_{\rm cr} = -m_1 \pm \sqrt{(m_1+\delta m)^2-\delta m^2}
\end{equation}
Still, \eqref{iah} are ambiguous, since it is not clear which of the inner horizon's segment is actually affected:
the outgoing ($v$-directed), the ingoing ($u$-directed) or both. To remove the ambiguity we have to use \eqref{eqmo}
\begin{equation}
\label{eqmo32}
\sqrt{\dot r^2+\left(1-\frac{m_1}{r}\right)^2}-\sigma_3\sqrt{\dot r^2+1-\frac{2(m_1+\delta m)}{r}+\frac{(m_1+\delta q)^2}{r^2}}=\frac{M}{r},
\end{equation}
here $\sigma_1=1$, since sector~1 is a part of the extremal RN. 
Again, the shell collapses from the infinity of the outer ${\cal R_+}$-region in sector~3, therefore initially $\sigma_3=1$.  
If the shell started to collapse with the initial proper velocity $v_0$, then \eqref{eqmo32} at infinity yields 
\begin{equation}
\delta m=M\sqrt{1+v^2_0}.
\end{equation} 
Thus, it is convenient to study \eqref{eqmo32} when parameters $M$, $v_0$ are fixed while $\delta q$ is varied.
Let us find from \eqref{eqmo32} the radius where $\sigma_3$ changes its sign
\begin{equation}
r_0=\frac{ r_{3+} r_{3-}+M^2-m^2}{ r_{3+} + r_{3-}-2m}
\end{equation} 
and plot it together with $ r_{3-}$ as functions of $\delta q$. The result is shown in Fig.~\ref{fig:4}. 
There are four intersection points of $r_0$, $r_{3-}$ at $(\delta q^{\pm}_1, m_1-M)$,  $(\delta q^{\pm}_2, m_1+M)$ with
\begin{eqnarray}
  \label{q12}
  \delta q^{\pm}_1 &=& -m_1 \pm \sqrt{(m_1+\delta m)^2-(\delta m+M)^2}\nonumber\\
  \delta q^{\pm}_2 &=& -m_1 \pm \sqrt{(m_1+\delta m)^2-(\delta m-M)^2}
\end{eqnarray}

The existence of two branches in \eqref{q12} and \eqref{qcrit} is a manifestation of the underlying $CP$ 
symmetry of the RN spacetime.
Indeed, the parity transformation in the RN  can be viewed  as mirroring  of diagram in Fig.~\ref{fig:0}(a) and mirroring 
and interchanging ${\cal R_{\pm}}$-regions in Fig.~\ref{fig:0}(b).
Then changing signs of all charges in the system to opposite, we ought to get the same dynamic. 
In the following, we choose the '+' branches of \eqref{q12} and \eqref{qcrit}.
Using  Fig.~\ref{fig:4} it is easy to trace possible types of the collapse. 

\begin{figure}[H]
\begin{center}
\begin{tabular}[c]{cc}
    \psfragscanon
    \psfrag{rp3}[c][c][1.2][0]{$r_{3+}$}
    \psfrag{rm2}[c][c][1.2][0]{$r_{2-}$}
    \psfrag{rp2}[c][c][1.2][0]{$r_{2+}$}
    \psfrag{a}[c][c][1.2][0]{$(a)$}
    \includegraphics[angle=0,width=0.25\textwidth]{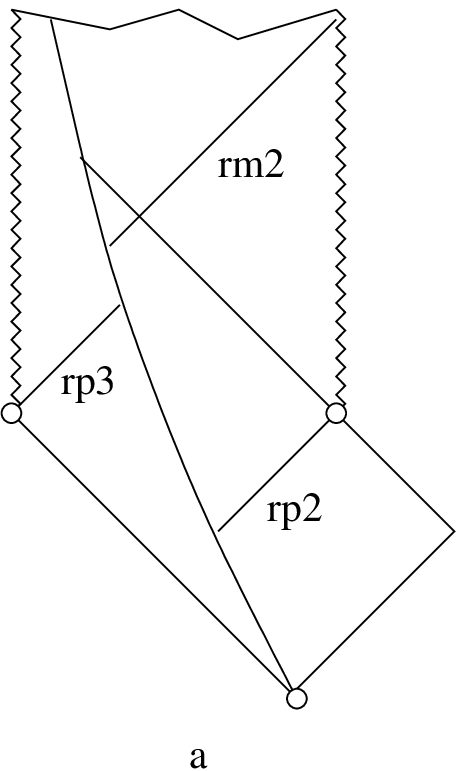} &
    \psfragscanon
    \psfrag{rp3}[c][c][1.2][0]{$r_{3+}$}
    \psfrag{rm2}[c][c][1.2][0]{$r_{2-}$}
    \psfrag{rp2}[c][c][1.2][0]{$r_{2+}$}
    \psfrag{b}[c][c][1.2][0]{$(b)$}
    \includegraphics[angle=0,width=0.25\textwidth]{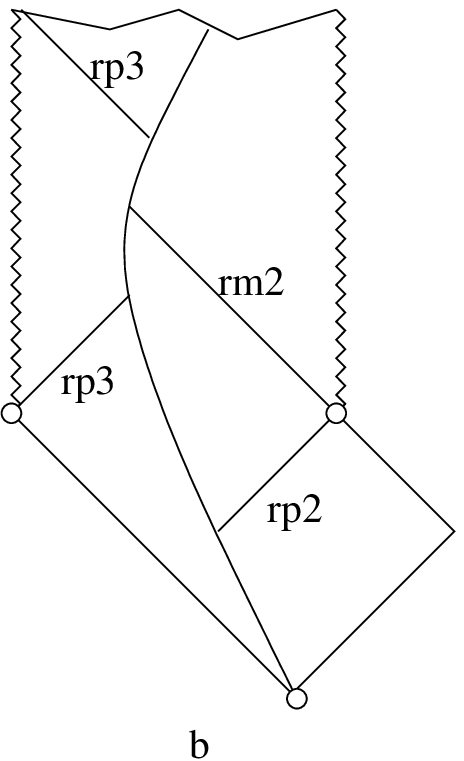}\\ \\
    \psfragscanon
    \psfrag{rp3}[c][c][1.2][0]{$r_{3+}$}
    \psfrag{rm2}[c][c][1.2][0]{$r_{2-}$}
    \psfrag{rp2}[c][c][1.2][0]{$r_{2+}$}
    \psfrag{c}[c][c][1.2][0]{$(c)$}
    \includegraphics[angle=0,width=0.25\textwidth]{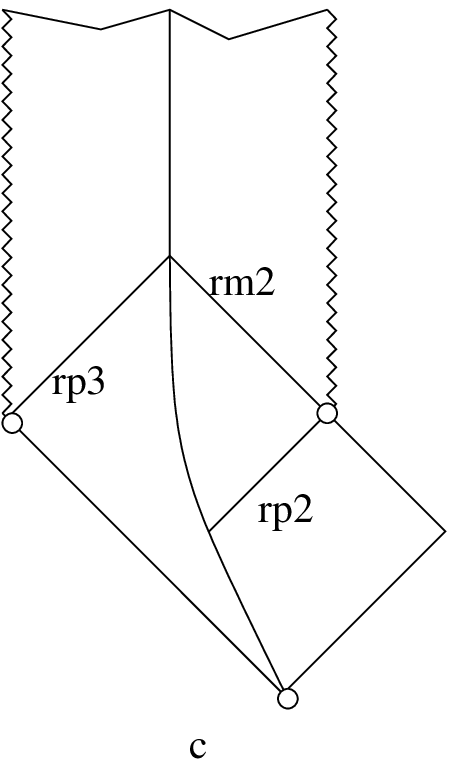} &
    \psfragscanon
    \psfrag{rp3}[c][c][1.2][0]{$r_{3+}$}
    \psfrag{rm2}[c][c][1.2][0]{$r_{2-}$}
    \psfrag{rp2}[c][c][1.2][0]{$r_{2+}$}
    \psfrag{d}[c][c][1.2][0]{$(d)$}
    \includegraphics[angle=0,width=0.25\textwidth]{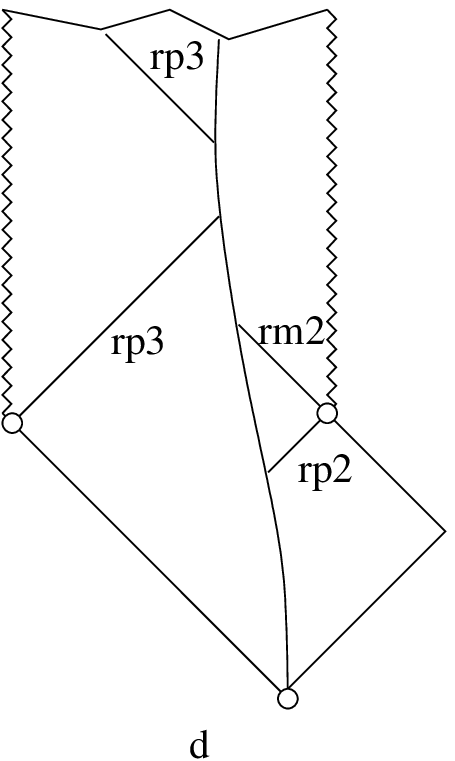}
\end{tabular}
\caption{Possible Carter-Penrose diagrams for the charged shell collapsing into the extremal RN black hole.}
\label{fig:5}
\end{center}
\end{figure}

When $\delta q < \delta q_1^+$, the shell undergoes collapse towards the inner ${\cal R_+}$-region as 
it is shown at Fig.~\ref{fig:5}(a).  However, in the interval $(\delta q_1^+, \delta q_2^+)$, $r_0$ 
lies in the ${\cal T_-}$-region and $\sigma_3$ can change its sign. This means that the shell can collapse towards 
the inner ${\cal R_-}$-region, see Fig.~\ref{fig:5}(b).
But before $\delta q^+_2$ is reached, the saturation of \eqref{iah} happens when $\delta q=\delta q^+_{\rm cr}$. The diagram for
this case is shown at Fig.~\ref{fig:5}(c). One can see that the crossing point lies exactly at the vertex of 
the Cauchy horizon common for both sectors~1,~3. Thus in the case  $r_{3-}= r_{1-}$, $\sigma_3$ always change its sign.
The case $\delta q > \delta q_{\rm cr}^+$ is described by diagram Fig.~\ref{fig:5}(d). 

It is interesting to note, that for $\delta q > \delta q_2^+$, $r_0$ lies in one of the inner ${\cal R}$-regions. 
This means, in fact, that $\sigma_3$ cannot be changed along the trajectory of the shell. 
However, let us calculate $\sigma_3$ at the turning point. 
For $\delta q > \delta q_2^+$ the turning point is given by the expression
\begin{equation}
\label{tp}
r_{\rm tp} = \frac{m_3(\delta q - M)+(\delta q^2- M^2)/2}{\delta m - M}.
\end{equation}
Then using \eqref{eqmo32} we get $\sigma_3=-1$. We can conclude that the shell with $\delta q > \delta q_2^+$
cannot collapses from infinity of the $\cal R_+$-region in sector~3. 
After these remarks we can return to the main subject.

To explore the spacetime after the crossing we have to use the extended DTR relation \eqref{edtr} with $F_3(r_c)\to 0$
along with \eqref{eqmo} immediately after the crossing 
\begin{eqnarray}
  \label{edtr_eom}
  \frac{\sigma_1\sigma_2\left(\sqrt{\dot r^2_b + F_1} - \sigma_1 \dot r_b \right)\left(\sqrt{\dot r^2_a + F_2} + \sigma_2 \dot r_a \right)}
  {\sigma_3\sigma_4\left( \sqrt{\dot r^2_a +F_4} - \sigma_4 \dot r_a\right)\left( \sqrt{\dot r^2_b +F_3} - \sigma_3 \dot r_b\right)} &=& 1,\nonumber\\
\sigma_4\sqrt{\dot r_a^2+F_4}-\sigma_2\sqrt{\dot r_a^2+F_2}&=&\frac{M}{r},
\end{eqnarray}
where unknowns are $\dot r_a$, $F_4$. 
From the above discussion we know that near the crossing point $\dot r_b<0$, and
$\sigma_1=1$, $\sigma_3=-1$. Also $F_2<0$ since the crossing point lies in $\cal T_-$-region
of sector~2.  Therefore we should successively solve \eqref{edtr_eom} for different values $\sigma_2$, $\sigma_4$.

\begin{description}

\item {\it Case $\sigma_2=-1$, $\sigma_4=1.$}
Solution of \eqref{edtr_eom} is
\begin{eqnarray}
\label{gsol1}
\dot r_a &=& - \frac{r_c(1+A)}{2M}F_2+\frac{M}{2r_c(1+A)},\\
F_4 &=& \frac{M^2}{r_c^2}+F_2-\frac{2M}{r_c}\sqrt{\dot r_a^2+F_2},\nonumber
\end{eqnarray}
with 
$$
A=\frac{\sqrt{\dot r^2_b + F_1} - \dot r_b}{ \sqrt{\dot r^2_b +F_3} + \dot r_b}.
$$
Since $F_3\to 0$ from below when $r\to r_c$, it follows that $A\to -\infty$ and consequently
\begin{eqnarray}
\label{sol1}
\dot r_a &=& -\infty\\
F_4 &=& -\infty\nonumber\quad\mbox{or}\quad m_4\to\infty.
\end{eqnarray}
The first expression of \eqref{sol1} means that the speed of the time-like shell tends to the speed of light after 
interaction with null shell. Informally, one could say that observers ``burns down'' while attempting 
to cross the Cauchy horizon.
The second expression of \eqref{sol1} is the mass inflation phenomenon.

\item {\it Case $\sigma_2=\sigma_4=-1.$} Solutions of \eqref{edtr_eom} coincide with \eqref{gsol1}, 
\eqref{sol1}.

\item {\it Case $\sigma_2=\sigma_4=1.$} In this situation  
\begin{eqnarray}
\label{gsol2}
\dot r_a &=& - \frac{r_c(1+A)}{2M}F_2+\frac{M}{2r_c(1+A)},\\
F_4 &=& \frac{M^2}{r_c^2}+F_2+\frac{2M}{r_c}\sqrt{\dot r_a^2+F_2},\nonumber
\end{eqnarray}
and when $A\to-\infty$ leads to a nonsense 
\begin{eqnarray}
\label{sol2}
\dot r_a &=& -\infty,\\
F_4 &=& \infty\nonumber\quad\mbox{or}\quad m_4\to-\infty.
\end{eqnarray}
This scenario is forbidden due to the following reasons. After the crossing, the timelike shell eventually enter into 
$\cal R_-$-region with respect to its exterior, thus $\sigma_2=-1$. This means that $\sigma_2$ should change its sign
in $\cal T_-$-region of sector~2. However in the limit $F_3\to 0$ this is impossible since the timelike shell is effectively
turns into a null shell.

\item {\it Case $\sigma_2=1$, $\sigma_4=-1.$} This possibility is forbidden as it follows from the second equation of 
\eqref{edtr_eom}.
\end{description} 

Thus, the picture is technically similar to the classical Israel-Poisson model. Moreover, the physics behind our model 
is also similar. On the one hand, trapped radiation is infinitely blue-shifted. On the other hand, generators 
of the Cauchy horizon have no conjugate points, thus no focussing. This two extremal conditions leads to an instability
of the horizon. Any matter that crosses it, will lead to catastrophic focussing at the inner apparent horizon and consequently
to the mass parameter blow up.

\begin{figure}[H]
  \begin{center}
 \psfragscanon
 \psfrag{r0}[c][c][0.9][90]{$r=0$}
 \psfrag{Ip}[c][c][0.9][0]{${\cal I}^{+}$}
 \psfrag{Im}[c][c][0.9][0]{${\cal I}^{-}$}
 \psfrag{ip}[c][c][0.9][0]{$i^+$}
 \psfrag{i0}[c][c][0.9][0]{$i^0$}
 \psfrag{Rp}[c][c][0.9][0]{${\cal R_+}$}
 \psfrag{Rm}[c][c][0.9][0]{${\cal R_-}$}
 \psfrag{Tm}[c][c][0.9][0]{${\cal T_-}$}
 \psfrag{A}[c][c][0.8][0]{$\rm 1$}
 \psfrag{B}[c][c][0.8][0]{$\rm 3$}
 \psfrag{C}[c][c][0.8][0]{$\rm 2$}
 \psfrag{D}[c][c][0.8][0]{$\rm 4$}
 \psfrag{a}[c][c][0.8][0]{$\rm 1'$}
 \psfrag{b}[c][c][0.8][0]{$\rm 3'$}
 \psfrag{c}[c][c][0.8][0]{$\rm 2'$}
 \psfrag{d}[c][c][0.8][0]{$\rm 4'$}
 \psfrag{rp}[c][c][0.9][0]{$r_{\rm 3+}$}
 \psfrag{rm}[c][c][0.9][0]{$r_{\rm 2-}$}
 \psfrag{eh}[c][c][0.9][0]{$r_{2+}$}
 \psfrag{ch}[c][c][0.9][0]{$r_{3-}$}
 \psfrag{ah}[c][c][0.9][0]{$r_{4-}$}
 \psfrag{u}[c][c][0.8][0]{$u$}
 \psfrag{v}[c][c][0.8][0]{$v$}
  \includegraphics[width=0.5\textwidth]{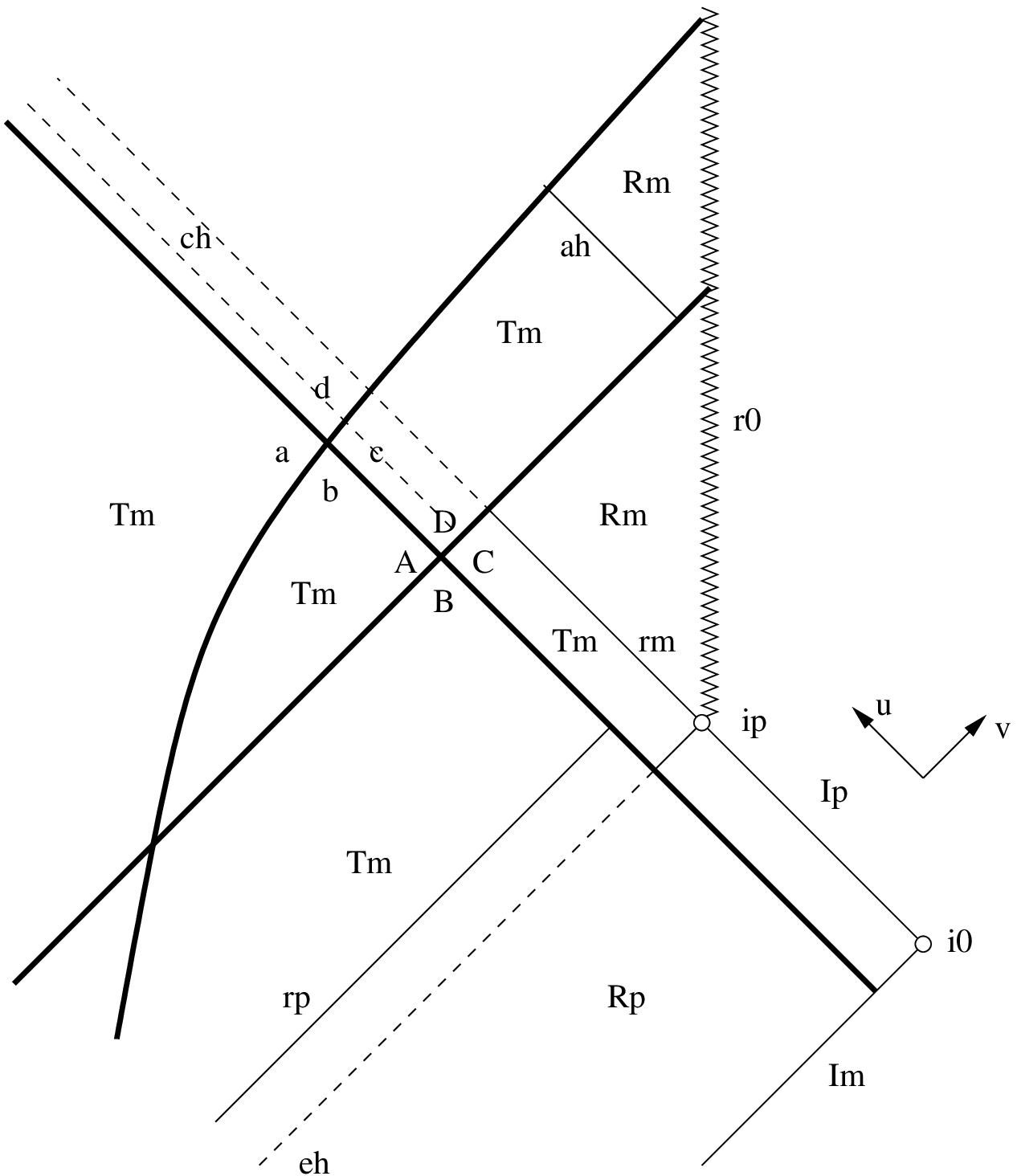}
\caption{}
\label{fig:6}
  \end{center}
\end{figure}

Another extension we are going to consider is a simplified model for collapsing interior of a charged black hole.
It is depicted in Fig.~\ref{fig:6}.
Two null shells of the original Israel-Poisson model describe infalling and backscattered radiation flows and 
a timelike shell describes massive (uncharged) particles falling towards the future Cauchy horizon. Two null shells cross
and create a region where the mass inflation takes place. Then the timelike shell interacts with the ingoing null shell
and enter into the mass inflation region (with respect to its exterior). The problem is to find the geometry between 
shells (sector 4'). This again can be done by using \eqref{edtr_eom} for the primed sectors. Since it is reasonable to consider
the interaction between the timelike shell and the outgoing null shell as a perturbation, then  $\sigma_{1'}=1$, $\sigma_{3'}=-1$
as in the previous model.  
Thus, for $\sigma_{4'}=1$, $\sigma_{2'}=-1$ and $\sigma_{4'}=-1$, $\sigma_{2'}=-1$, the unknowns $\dot r_a, F_{4'}$ is defined by \eqref{gsol1}.
At this time however, $A$ is finite since $m_{3'}=m_{1}\neq m_{3}$. Also, by construction, sector 4 of the original Israel-Poisson 
model coincides with sector 2', thus $m_{2'}\to\infty$. Therefore \eqref{gsol1} gives 
$$
\dot r_a \to -\infty,\quad \mbox{when} \quad F_{2'}\to -\infty.
$$
Thus the timelike shell ``burns down'' and eventually hits the singularity.

On the other hand, eliminating $\dot r_a$ form the second equation \eqref{gsol1} we obtain
$$
F_{4'}=\frac{M^2}{r_c}+F_{2'}-\frac{2M}{r_c}|F_{2'}|\sqrt{\left[\frac{r_c(1+A)}{2M}+\frac{M}{2r_c(1+A)F_{2'}}\right]^2+\frac{1}{F_{2'}}}.
$$
In general $F_{4'}\to-\infty$ when $F_{2'}\to-\infty$, however $F_{4'}$ can be made finite and positive in this limit when
\begin{equation}
\label{ft}
r_c=\frac{2M}{(1+A)^2}
\end{equation}
Thus, it is possible to ``switch off'' the mass inflation by carefully choosing $\dot r_b$ and $M$ and open the RN wormhole again.
It is not clear if a similar effect will persist in more advanced model without thin shells. We believe it is hardly possible,
since \eqref{ft} represents rather specific fine tuning and we do not see any deeper physical explanation for this relation. 
  
The remained cases $\sigma_{2'}=\sigma_{4'}=1$ and $\sigma_{4'}=-1,\sigma_{2'}=1$ are forbidden due to the same reasons
as in the previous model. 

\section{Conclusions}
\label{sec:2}
The main objective of the present paper was to show some application examples of the extended DTR \eqref{edtr}. 
In particular, we considered
two extensions of the Israel-Poisson model. In the first model a charged timelike shell crosses the null shell propagated 
along the future Cauchy horizon. This extension can be interpreted as the   
simplest model with backreaction for observers crossing the future Cauchy horizon with trapped radiation. 
Then using the extended DTR we have shown that in our model observers turn into light 
while crossing the Cauchy horizon. Moreover the mass inflation also takes place.  

Another extension describes a simplified model for collapsing interior of charged black hole. Here the timelike shell crosses
the Cauchy horizon {\it after} the mass inflation was triggered. It appears that there is a combination of parameters 
which allows to ``switch~off'' the mass inflation.

With regard to other applications of the extended DTR,
there is a field of gravitational physics where this relation can be especially useful. This is the physics of wormholes.
In particular, a large class of traversable wormholes was constructed by gluing different spherically-symmetric spacetimes along 
timelike shells \cite{Visser}.  Also, one can construct traversable wormholes by using thin null shells with negative energy density. See for example \cite{Hazboun}. In this respect, the extended DTR could be used to test stability and traversability properties of such wormholes when backreaction is taken into account.

\section*{Acknowledgements}
I am grateful to Victor Berezin for his invaluable comments. This work is partially supported by RFBR grant 13-02-00257a.

\end{document}